\documentclass[epsfig,graphics,twocolumn,showpacs,floatfix,mathbbm,prl,superscriptaddress]{revtex4}
\usepackage{amsmath,amssymb,graphicx,color}
\usepackage{float}
\usepackage{subfigure}
\newcommand{\PrYSO}{Pr$^{3+}$:Y$_2$Si{O$_5$}}
\newcommand{\YSO}{Y$_2$Si{O$_5$}}

\newcommand{\NdYVO}{Nd:YV{O$_4$}}
\newcommand{\mumu}{$\mu\mathrm{m}$}
\newcommand{\mumus}{$\mu\mathrm{m/s}$}

\begin{document}

\title{An Integrated Optical Memory based on Laser Written Waveguides}
\pacs{42.25.Kb, 42.50.Gy, 42.50.Md, 42.82.Bq}

\author{Giacomo Corrielli}
\altaffiliation{These authors contributed equally to this paper}
\affiliation{Istituto di Fotonica e Nanotecnologie - Consiglio Nazionale delle Ricerche and Dipartimento di Fisica - Politecnico di Milano, 5 P.zza Leonardo da Vinci 32, 20133 Milano, Italia}
\author{Alessandro Seri}
\altaffiliation{These authors contributed equally to this paper}
\affiliation{ICFO-Institut de Ciencies Fotoniques, The Barcelona Institute of Technology, 08860 Castelldefels (Barcelona), Spain}
\author{Margherita Mazzera}
\email{margherita.mazzera@icfo.es}
\affiliation{ICFO-Institut de Ciencies Fotoniques, The Barcelona Institute of Technology, 08860 Castelldefels (Barcelona), Spain}
\author{Roberto Osellame}
\affiliation{Istituto di Fotonica e Nanotecnologie - Consiglio Nazionale delle Ricerche and Dipartimento di Fisica - Politecnico di Milano, 5 P.zza Leonardo da Vinci 32, 20133 Milano, Italia}
\author{Hugues de Riedmatten}
\affiliation{ICFO-Institut de Ciencies Fotoniques, The Barcelona Institute of Technology, 08860 Castelldefels (Barcelona), Spain}
\affiliation{ICREA-Instituci\'{o} Catalana de Recerca i Estudis Avan\c cats, 08015 Barcelona, Spain}

\date{\today}
\begin{abstract} 

We propose and demonstrate a new physical platform for the realization of integrated photonic memories, based on laser-written waveguides in rare-earth doped crystals. Using femto-second laser micromachining, we fabricate waveguides in \PrYSO \, crystal. We demonstrate that the waveguide inscription does not affect the coherence properties of the material and that the light confinement in the waveguide increases the interaction with the active ions by a factor 6. We also demonstrate that, analogously to the bulk crystals, we can operate the optical pumping protocols necessary to prepare the population in atomic frequency combs, that we use to demonstrate light storage in excited and spin states of the Praseodymium ions. Our results represent the first realization of laser written waveguides in a \PrYSO \, crystal and the first implementation of an integrated on-demand spin wave optical memory. They open new perspectives for integrated quantum memories. 
\end{abstract}

\maketitle
\section{Introduction}

Quantum memories (QMs) are important devices in quantum information science. They provide an interface between flying and stationary quantum bits and are at the heart of several applications including quantum information networks \cite{Kimble2008}, quantum repeaters \cite{Sangouard2011}, linear optics quantum computing \cite{Knill2001}, and  multi-photon quantum light sources \cite{Nunn2013}. Several QM demonstrations have been reported recently in atomic and solid state systems \cite{Lvovsky2009a, Sangouard2011,Bussieres2013}. In order to progress towards large scale quantum information architectures involving several QMs, it is important to rely on devices  that can be easily duplicated and integrated. This would facilitate the scalability and the realization of complex optical circuitry involving QMs. Furthermore, the tight light confinement achieved in waveguide structures would lead to a strong enhancement of the light matter interaction. Finally, the integration of QMs with other required devices, such as quantum light sources and single photon detectors, would be greatly simplified. Solid state systems are well suited for the exploration of integrated QMs, and in particular rare-earth doped solids showed promising properties with bulk crystals \cite{Bussieres2013, Sangouard2011}.

Two approaches have been explored towards integrated rare-earth QMs so far. The first one is to integrate rare earths in already available systems for waveguides \cite{Staudt2007, Staudt2007a}.  Quantum light storage has been  performed in a Ti$^{4+}$:Tm$^{3+}$:LiNbO$_3$ waveguide fabricated by Ti$^{4+}$ in-diffusion \cite{Saglamyurek2011}, where also more recently an integrated processor has been implemented \cite{Saglamyurek2014}. These demonstrations were  limited to the mapping of light field into optical atomic excitations. Moveover, the LiNbO$_3$ matrix, although suitable for waveguide fabrication, degrades the coherence properties of the rare-earth ions, which then limits the storage performances, e.g. in terms of storage times \cite{Saglamyurek2011}. The second approach consists in realizing waveguides in crystals already used for demonstrating bulk QM and suitable for long term storage of quantum information, e.g. \YSO. Photonic crystal waveguides and cavities have been fabricated in {Nd$^{3+}$:Y$_2$Si{O$_5$}} and  {Er$^{3+}$:Y$_2$Si{O$_5$}}, using focused ion beam milling \cite{Zhong2015}. It was shown that the coherence properties of the rare-earth ions are preserved during the fabrication process, which is very promising for quantum applications. This fabrication technique is however challenging to extend to mm long waveguides, which may be useful to achieve high storage and retrieval efficiencies. Also, the ions used so far possess only two-fold ground states, which restricts  the storage of light to the excited state, strongly limiting the achievable storage times. In order to attain spin-state storage, which enables long term storage and on demand-read-out, ions with a three-fold ground state, such as  Eu$^{3+}$ or Pr$^{3+}$, should be used \cite{Afzelius2010, Gundogan2013,Jobez2015,Gundogan2015}. \PrYSO \, crystals are currently one of the best systems for quantum memory applications. Very efficient storage of weak coherent states \cite{Hedges2010} has been demonstrated in this material, as well as the longest storage time ever demonstrated in any system, in the order of 1 minute for classical images \cite{Heinze2013}. Recently the first  demonstration of an on-demand quantum memory for time-bin qubits \cite{Gundogan2015} has been reported using this crystal. The coherence properties of devices based on TeO$_2$ slab waveguides deposited on a \PrYSO\, crystal have been measured to be consistent with those of bulk ions \cite{Marzban2015}, which is promising for the implementation of integrated rare-earth based quantum devices. Nevertheless, in such a system the active ions are only evanescently coupled to the waveguide.

Here, we propose an alternative way to fabricate waveguides in \PrYSO \, using femtosecond laser micromachining (FLM), where the active ions are directly coupled to the light. FLM demonstrated in the past two decades to be a very powerful technology for the direct inscription of high quality optical waveguides in the bulk of both amorphous and crystalline transparent substrates \cite{davis1996writing,Valle2009,Gorelik2003,Thomson2006}. Several complex integrated photonic devices have been developed with this technique, ranging from all optical routers \cite{keil2011all} and power dividers \cite{liu2005directly}, to modulators \cite{liao2008electro}, and frequency converters \cite{osellame2007femtosecond,burghoff2006efficient}. Moreover, it has been shown that laser written waveguide circuits in glass are suitable for supporting the propagation of photonic qubits \cite{marshall2009laser}, and represent a promising platform for the development of the rapidly growing field of integrated quantum photonics \cite{vest2015design,crespi2013integrated,tillmann2013experimental}. The class of materials where waveguide writing with FLM has been demonstrated includes several rare-earth doped crystals, e.g. Nd:YAG, Yb:YAG, \NdYVO \,, and Pr:YLF among others, with applications mainly oriented towards the realization of integrated laser sources \cite{calmano2010nd,calmano2011diode,tan2011simultaneous,muller2012femtosecond}. At the best of our knowledge, FLM in \PrYSO \, crystals has never been reported in the literature before, despite it would allow to take advantage of the direct access to the active ions and the exceptional performances as light storage medium. 

In this paper, we demonstrate that the coherence properties of Pr$^{3+}$ ions in fs-laser written waveguides are not affected by the fabrication procedure. We show that the light confinement in the waveguide improves significantly the interaction with the active ions. 
Finally, we perform storage and retrieval of light pulses using the atomic frequency comb (AFC) protocol, both in the excited state and spin-states, demonstrating the first on-demand integrated optical memory.

\section{Experimental setup and fabrication}
The substrate used is a \PrYSO \, bulk crystal (Scientific Material), $10 \,\mathrm{mm}$-long (along the crystal $b$-axis) and with a concentration of active ions of $ 0.05\,\%$. Optical waveguides are fabricated by FLM adopting the so-called type II configuration \cite{chen2014optical}, where the fs-laser irradiation is used to directly inscribe into the substrate two closely spaced damage tracks, in correspondence of which the material locally expands and becomes amorphous. This procedure gives rise to the formation of a stress field in their proximity, which, in turn, causes a material refractive index alteration. By tailoring properly the irradiation parameters and geometry, it is possible, with this method, to obtain a light guiding region, with positive refractive index change, localized between the two tracks. It is worth to highlight that in type II waveguides the core region (where most of the light remains confined) is only marginally affected by the fabrication process, hence preserving all its bulk properties. Single mode waveguides for 606 nm light have been fabricated accordingly, by inscribing pairs of damage tracks with a reciprocal distance of 25 \mumu \, and buried 100 \mumu \, beneath the sample top surface. Each track is fabricated by focusing a femtosecond laser beam (a home made Yb:KYW oscillator, $\lambda$=1030 nm, pulse duration $\tau$=300 fs \cite{Killi2004}) inside the crystal volume, with optimized irradiation parameters: energy per pulse of 570 nJ, repetition rate of 20 kHz, uniform sample translation along the $b$-axis at the speed of 57 \mumus. A microscope objective with 50x magnification and 0.6 numerical aperture is used as focusing optics. Note that, for higher laser repetition rates and equal energy deposited per unit volume (at a fixed pulse energy) we observed the formation of periodic disruptions along the damage tracks \cite{richter2013formation}, which degrade the waveguide homogeneity and increase significantly the waveguide propagation losses.
During the fabrication, the crystal orientation is such that the fs-pulsed laser beam is parallel to the $D_1$-axis. The resulting waveguides support only one polarization mode, with the electric field oscillating parallel to the $D_2$-axis, which is compatible with the memory protocols, as it is the one that maximizes the interaction between the Pr$^{3+}$ ions and the light field resonant to their $^3 H_4(0) \rightarrow ^1 D_2(0)$ optical transition at $606\, \mathrm{nm}$ \cite{Sun2005}.

\begin{figure}[ht]
\centering

{\includegraphics[width=1\linewidth]{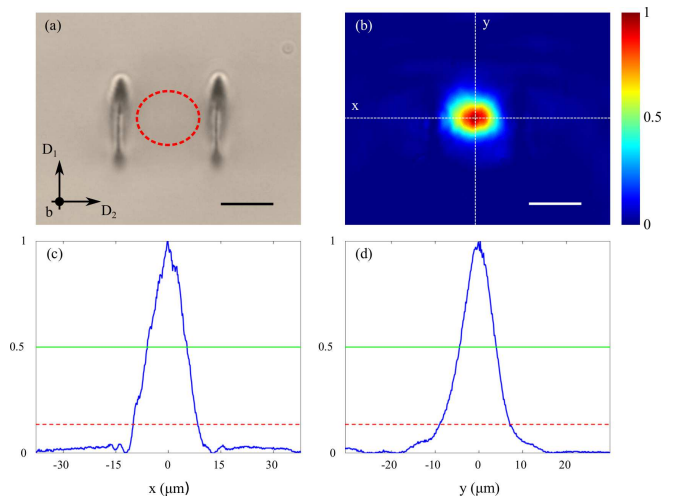}}
\caption{(a) Microscope picture of the waveguide cross section. The distance between the damage tracks is 25 \mumu . The red-dashed ellipse indicates the $e^{-2}$ contour of the guided mode. Scale bar is 15 \mumu. (b) CCD-acquired near-field intensity profile of the guided mode. Scale bar is 15 \mumu . (c),(d) Normalized intensity profiles of the waveguide mode along the x and y sections indicated in panel (b). The resulting full widths at half maximum ($FWHM$) are $\Delta_x = 11.3\, \mu \mathrm{m}$ and $\Delta_y = 8.6\, \mu \mathrm{m}$ (green solid level). The measured $e^{-2}$ diameters are $\tau_x = 18.5\, \mu \mathrm{m}$ and $\tau_y = 15.8\, \mu \mathrm{m}$ (red dashed level).}
\label{modes}
\end{figure}

Our laser source at $606\, \mathrm{nm}$, a second harmonic generation laser (Toptica, DL-SHG-pro), is stabilized in frequency by the Pound-Drever-Hall technique to a Fabry-Perot cavity hosted in a home made vacuum chamber \cite{Rielander2014}. 
Both the amplitude and the frequency are modulated with a double pass acousto-optic modulator driven by an arbitrary waveform generator (Signadyne).
We use the same beam for both memory preparation and input light pulses. After the modulating AOM, the beam is steered to a separated optical table hosting the cryostat (closed cycle cryocooler, Oxford Instruments) where the waveguide sample is fixed and maintained at a temperature of about $3\, \mathrm{K}$. 
The coupling of the light into the waveguide is done by means of an external $75\, \mathrm{mm}$ focal length lens assembled in a translation stage. The final $e^{-2}$ diameter of the beam at the waveguide input facet is $28.3\, \mu \mathrm{m}$. 
After the waveguide, the light is recollected with a $100\, \mathrm{mm}$ focal length lens and sent to a photodetector (or CCD camera) placed at a distance of about $180\, \mathrm{cm}$ from the collecting lens.

\section{Optical measurements}

Figure \ref{modes}(a) shows a microscope picture of the waveguide cross section, in which the two damages are clearly visible. The mode supported by the waveguide, when it is coupled with 606 nm light polarized parallel to the \YSO \, crystalline $D_2$-axis, is reported in Fig. \ref{modes}(b).
From the horizontal and vertical intensity profiles, shown in panels (c) and (d), respectively, we estimate the full widths at half maximum ($FWHM$) of the guided mode to be $\Delta_x = 11.3\, \mu \mathrm{m}$ and $\Delta_y = 8.6\, \mu \mathrm{m}$. The transmission of the light through the optical waveguide is about $\eta_T = 50\,\%$, which includes coupling mode mismatch and waveguide propagation loss. Fresnel losses at the waveguide input and output facets are quenched by a specific anti-reflection coating. By rotating the polarization of $90\, ^{\circ}$ the spot almost disappears, which confirms that the light is guided through the waveguide. As a matter of fact, in a bulk crystal the polarization perpendicular to the $D_2$-axis would only weakly interact with the Pr$^{3+}$ ions, thus experiencing a higher transmission through the sample.

\begin{figure}[h]
\centerline{\includegraphics[width=1\columnwidth]{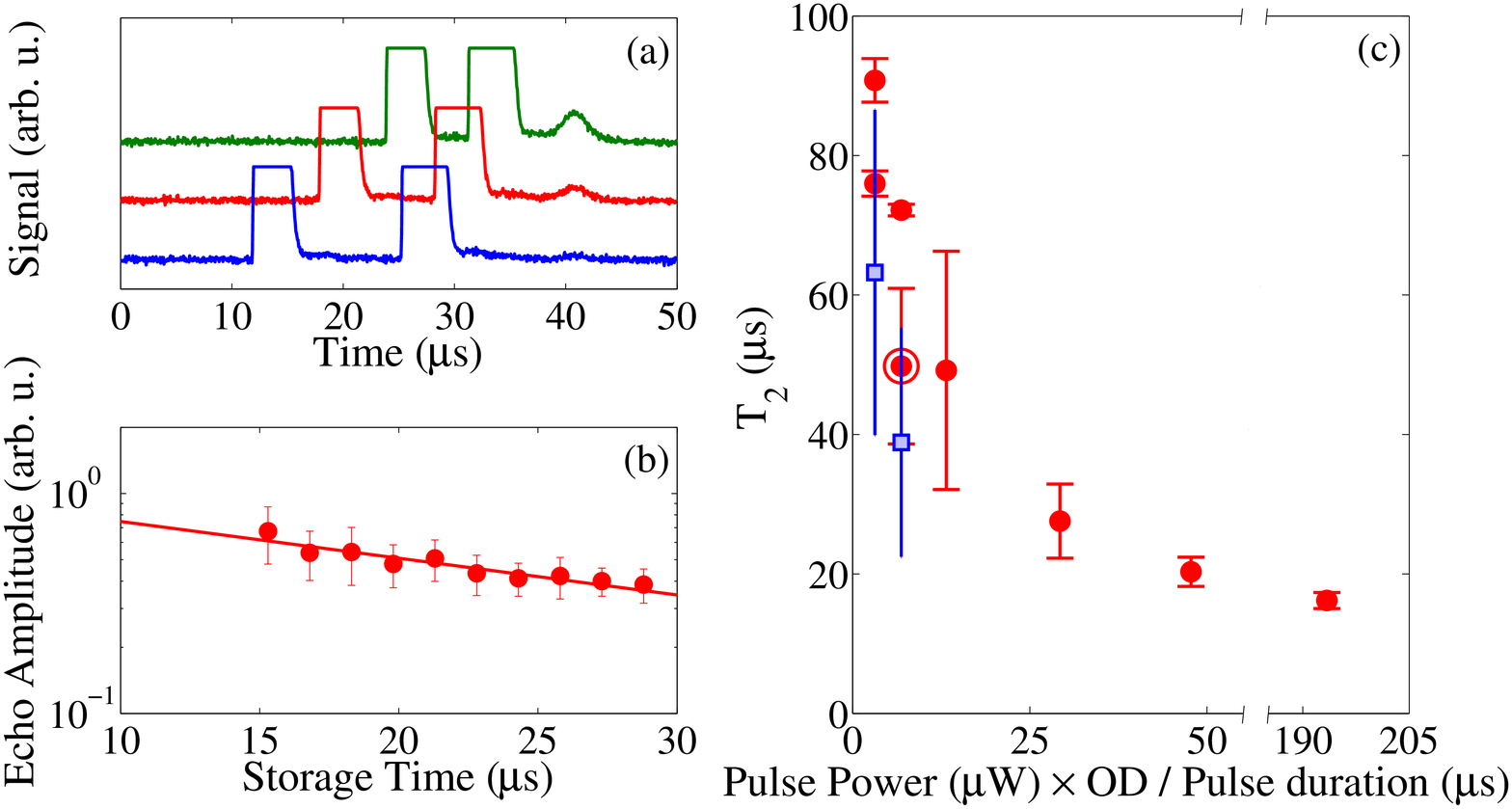}}
\caption{(a) Pulse sequence for the two-pulse photon echo experiment. (b) Echo efficiency as a function of the time delay $2\tau$. The fitting of the experimental data to an exponential decay is also shown (solid line). For this set of data an optical coherence time $T_2 = 49.9\, \mu s$ is extrapolated. (c) Optical coherence times $T_2$ measured in the waveguide (circles) and in an equivalent bulk sample (squares) as a function of the quantity $\frac{P_p \times OD}{t_p}$. The circled data point in panel (c) refers to the decay shown in panel (b). }
\label{2PE}
\end{figure}

\begin{figure}
\centering\includegraphics[width=1\columnwidth]{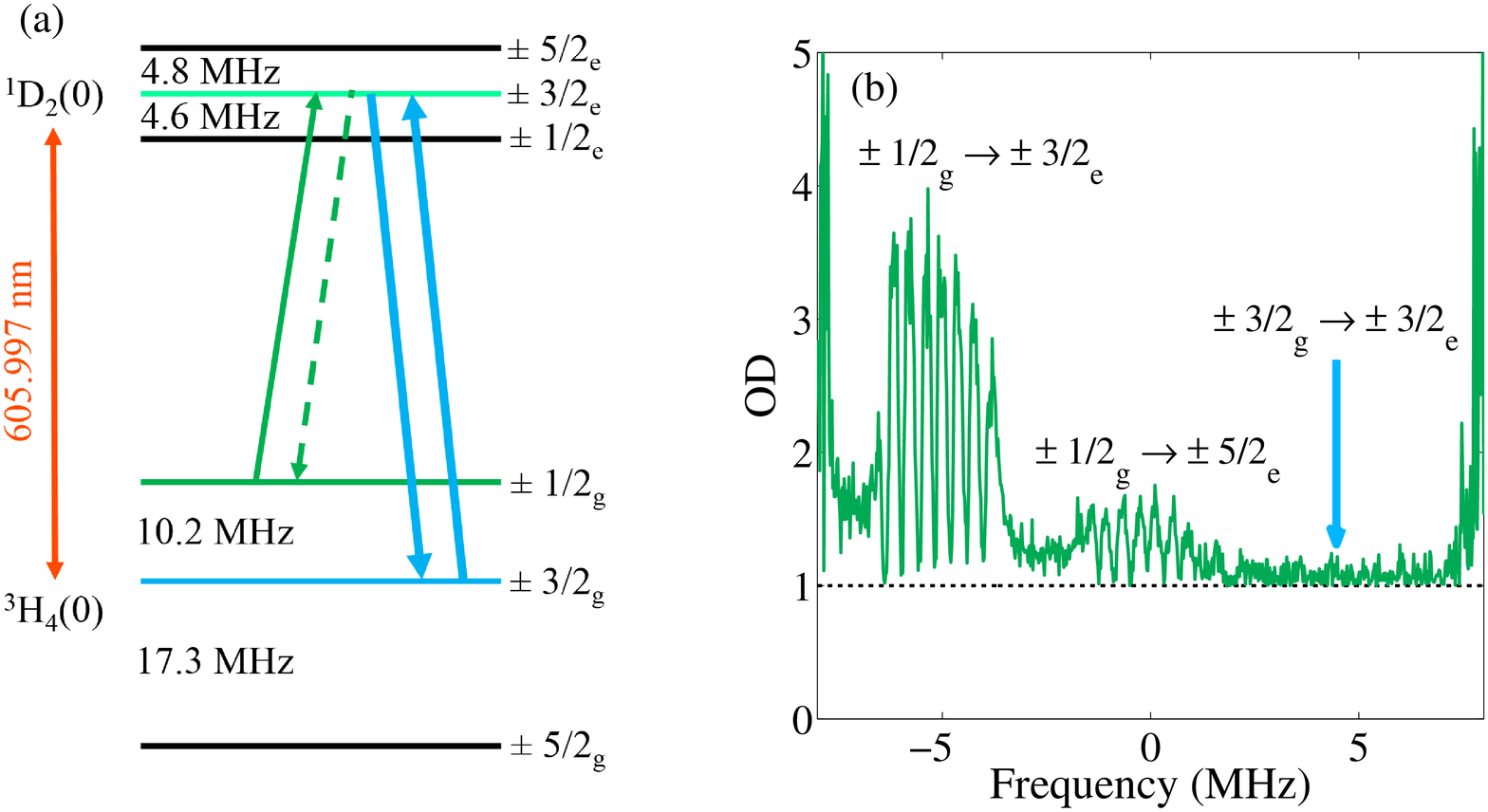}
\caption{(a) Energy level scheme reporting the hyperfine separation of the lowest electronic sublevels (0) of the $^3 H_4$ ground and $^1 D_2$ excited manifolds of Pr$^{3+}$ in \YSO. The $\Lambda$ scheme chosen for the storage is indicated by arrows. (b) Example of an atomic frequency comb (AFC) prepared at the frequency of the  ${1}/{2}_\textrm{g} \rightarrow {3}/{2}_\textrm{e}$ transition. } 
\label{AFC}
\end{figure}

To probe the coherence properties of the material after the waveguide fabrication, we perform two-pulse photon echo experiments and measure the optical coherence time $T_2$. The pulse sequence used is shown in Fig. \ref{2PE}(a). It starts with an input pulse resonant with the ${1}/{2}_\textrm{g} \rightarrow {3}/{2}_\textrm{e}$ transition between the ground $^3H_4(0)$ and excited $^1D_2(0)$ manifold of Pr$^{3+}$ (see level scheme in Fig. \ref{AFC} (a)),  which creates a coherent superposition of atomic excitations. A subsequent $\pi$-pulse after a time $\tau$ is used to refocus the dipoles associated to the collective excitation and induces thus the so called photon-echo after a time $2\tau$. When the time delay $\tau$ is increased, the amplitude of the echo is reduced due to the atomic decoherence. We thus estimate the optical coherence time $T_2$ from the decay of the echo efficiency (Fig. \ref{2PE}(b)). We measure $T_2$  in single class absorption features of different optical depths ($OD$) on the ${1}/{2}_\textrm{g} \rightarrow {3}/{2}_\textrm{e}$ transition \cite{Gundogan2015} prepared by optical pumping  and by modulating the pulse power $P_p$. To ensure efficient population transfer even when decreasing the pulse power, the duration of the pulses $t_p$ is simultaneously increased, thus reducing their spectral bandwidth. Figure \ref{2PE}(c) summarizes the $T_2$ value obtained as a function of the quantity $\frac{P_p \times OD }{t_p}$, which gives a measure of the excitation intensity (full red circles). As expected, the less ions are excited by the pulses, the longer coherence times are obtained due to the suppression of instantaneous spectral diffusion effects \cite{Equall1995}. It is worth noting that the obtained values agree with those, also reported in Fig. \ref{2PE} (open blue squares), measured when the laser beam is shifted towards the center of the \PrYSO \, bulk, far from the laser written waveguide, but maintaining the same focusing conditions. This agreement demonstrates that the micromachining procedure does not affect the coherence properties of the material in the spatial mode where the light is guided.
We also verified that the inhomogeneous broadening of the optical transition was preserved after the waveguide fabrication. The $FWHM$ of the transmission of the $^3H_4(0) \rightarrow \, ^1D_2(0)$ optical transition of Pr$^{3+}$ was measured to be about $20\,\mathrm{GHz}$, compatible with that measured in the same bulk sample.

\begin{figure}
\centering\includegraphics[width=1\columnwidth]{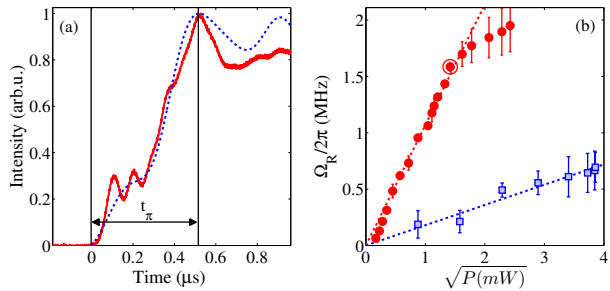}
\caption{ (a) Intensity of a long light pulse, $P_p \approx 2\, \mathrm{mW}$, transmitted by a single class absorption feature with optical depth $OD = 2.35$. The Rabi frequency $\Omega_R ^{WG} \approx 2\pi \cdot 1.6 \, \mathrm{MHz}$ is estimated from the rising time $t_{\pi}$ (indicated in the figure with vertical solid lines). (b) Rabi frequency as a function of the pulse power as calculated from optical nutation measurements performed on the waveguide (circles) and on the bulk (squares). The circled data point in panel (b) refers to the pulse reported in panel (a).}
\label{rabi}
\end{figure}

The strength of the interaction between the light and the active ions is associated to the Rabi frequency $\Omega_R$ of the optical transition, which we measure by means of optical nutation \cite{Sun2000}. We prepare a single class $2.5\,\mathrm{MHz}$ wide absorption feature at the frequency of the optical transition to probe (${1}/{2}_\textrm{g} \rightarrow {3}/{2}_\textrm{e}$ between the ground $^3H_4(0)$ and excited $^1D_2(0)$ manifold). We then send long square resonant pulses and measure the frequency of the coherent oscillation of the atoms under the light field. For inhomogeneously broadened transition and Gaussian beam profiles, the power transmitted is expected to oscillate as $J_1(\Omega_R t)/\Omega_R t$ \cite{Sun2000}. Fig. \ref{rabi}(a) reports an example, where the long pulse, of power $P_p \approx 2\, \mathrm{mW}$, is collected at the photodiode after the transmission through the optical waveguide (solid curve). The pulse exhibits a fast oscillation with frequency about $10.2\, \mathrm{MHz}$, i.e. the separation between the  ${1}/{2}_\textrm{g}$ and ${3}/{2}_\textrm{g}$ ground states. We attribute the oscillation to beatings between the two transitions ${1}/{2}_\textrm{g} - {3}/{2}_\textrm{e}$ and ${3}/{2}_\textrm{g} - {3}/{2}_\textrm{e}$ which might be simultaneously excited by the pulse due to imperfect optical pumping. The Rabi frequency is calculated in a trace corrected by the fast oscillation (gray dashed curve in Fig. \ref{rabi}(a)) from the population inversion time, $t_{\pi}$ indicated with the dotted vertical lines, as $\Omega_R \cdot t_{\pi}= 5.1$. The Rabi frequencies measured for different input light powers coupled in the waveguide are shown in Fig. \ref{rabi}(b) as a function of the square root of the input power. For input powers up to $\approx 2.6\,\mathrm{mW}$ the Rabi frequency scales as the square root of the power, as expected, but at higher powers the slope changes as if a saturation occurred. This saturation may be due to the fact that the Rabi frequency becomes comparable to the absorption feature width. Further measurements are necessary to confirm this hypothesis.
When moving the beam into the bulk (black squares), the Rabi frequency increases linearly with the square root of the power with a much lower slope than that observed in the waveguide (the maximum being $\Omega_R ^b \approx 2\pi  \cdot 690 \, \mathrm{kHz}$ at $15\,\mathrm{mW}$). For input powers at which the behavior is linear, the waveguide features light-ion interaction strengths higher than in the bulk by about a factor 6, due to the efficient light confinement. This value agrees with that theoretically obtained ($\approx 5.7$) by comparing the average Rabi frequency of a beam propagating along the bulk (refractive index $n = 1.8$) focused at the input facet with our measured waist of $14\,\mu \mathrm{m}$ and of a beam confined in the waveguide with $\eta_T = 50\,\%$. Note that this factor could be readily increased by optimizing the mode matching and thus decreasing the insertion losses in the present waveguide.

\section{Light storage}

Finally, we test our system as a storage device. The chosen storage protocol is the atomic frequency comb (AFC) \cite{afzelius2009}. It consists in tailoring a periodic structure, with $\Delta$ the periodicity, in a transparency window created within the inhomogeneously broadened optical transition $^3 H_4(0) \rightarrow ^1 D_2(0)$ of  Pr$^{3+}$. A light pulse absorbed by the structure creates a collective optical excitation which, after an initial dephasing, rephases at a time $\tau = \frac{1}{\Delta}$,  giving rise to a photon-echo like collective emission in the forward mode. Before the emission of the AFC echo, the collective optical excitation can be mapped into a spin excitation (spin wave) by applying a strong transfer pulse, thus effectively stopping the atomic dipole dephasing. The necessary requirement for the spin wave storage is to have a system with  three ground states, one where the AFC is prepared, one empty where the excitations are transferred, and one to use as auxiliary state for the AFC preparation. A second transfer pulse is applied to retrieve the stored pulse. 
Details about the procedure to create the AFC are provided in \cite{Gundogan2015}.
Fig. \ref{AFC}(a) shows the relevant level scheme  of the  Pr$^{3+}$ involved in the storage with the arrows indicating the frequencies of the input and transfer pulses. In the present experiment, the ground states $1/2_g$, $3/2_g$, and $5/2_g$, are chosen as the AFC preparation, the spin-wave storage, and the auxiliary state, respectively. An example of AFC structure with periodicity $\Delta = 400\, \mathrm{kHz}$ obtained by optical pumping in the waveguide at the frequency of the ${1}/{2}_\textrm{g} \rightarrow {3}/{2}_\textrm{e}$ transition is shown in Fig. \ref{AFC}(b). 

\begin{figure}
\centering\includegraphics[width=1\columnwidth]{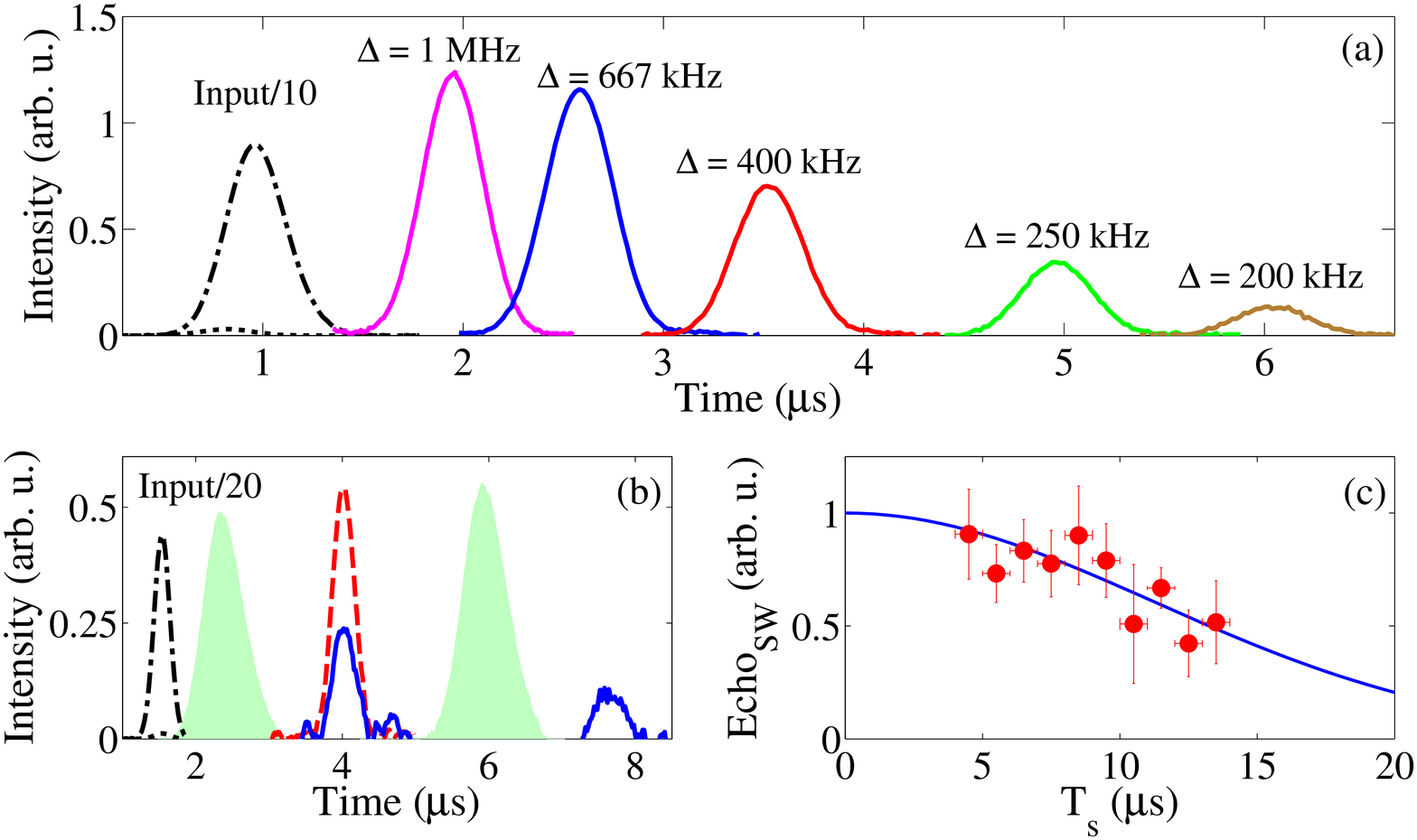}
\caption{Light storage experiments using the AFC protocol. (a) Storage in excited state. Dash-dotted line: $345\,\mathrm{ns}$ long input pulse transmitted through a transparency window (divided by 10 for clarity), in absence of AFC, polarized parallel to the \YSO\, crystalline $D_2$-axis. Dotted line:  input pulse polarized perpendicular to the  $D_2$-axis. Solid lines: AFC echoes for different AFC spacings $\Delta$. (b) Spin-wave storage. Black dash-dotted curve: similar as in (a) but $260\,\mathrm{ns}$ long (divided by 20 for clarity). Red dashed line: AFC echo in absence of control pulse, at a storage time of $\tau = 2.5\, \mu \mathrm{s}$. The internal efficiency is $\eta_{AFC}= 8.3 \%$. The green plain pulses represent the control pulses, as detected before the sample by a reference detector. The blue solid line is the output when control pulses are applied , with a time difference of $T_S=3.6 \mu \mathrm{s}$. The AFC echo is partially suppressed and a spin-wave echo appears $6.1 \mu \mathrm{s}$ after the input. The internal efficiency of the spin-wave echo is 2 $\%$. The technical noise due to the detector has been subtracted from the spin-wave storage trace. (c) Normalized spin-wave echo intensities as a function of the storage time $T_S$. The experimental data (dots) are fitted to a Gaussian decay to account for the inhomogeneous spin-broadening, from which we obtain $\gamma_{inh} = (23.6\pm 7.7) \,\mathrm{kHz}$.} 
\label{SWS}
\end{figure}

The results of the AFC storage experiments performed with the waveguide are shown in Fig. \ref{SWS}. In Fig. \ref{SWS}(a), a Gaussian light pulse ($FWHM = 345\, \mathrm{ns}$) linearly polarized parallel to the $D_2$-axis of \YSO \, is first sent through a transparency window, about $18\,\mathrm{MHz}$-wide, prepared in the Pr$^{3+}$ absorption line (black dash-dotted curves). When the polarization is rotated by $90\, ^{\circ}$ (black dashed curves), the light pulse at the output of the waveguide is strongly suppressed (see panel (a)), as the waveguide efficiently supports only the polarization parallel to the $D_2$ axis. 
The solid lines in Fig. \ref{SWS}(a) correspond to the AFC echo for different AFC spacing $\Delta$.  The maximal internal storage and retrieval efficiency $\eta_{AFC}$ (calculated as the ratio between the AFC echo and the input pulse transmitted through a transparency window) achieved is 14.6 $\%$ for a storage time of  $1.5 \,\mu \mathrm{s}$. The efficiency decreases when the storage time is increased, due to a reduction of comb quality and finesse \cite{afzelius2009,Rielander2014}. The device efficiency $\eta_{AFC}^{d}$, defined as the AFC echo after the crystal divided by the input pulse before the crystal, can be estimated by multiplying  $\eta_{AFC}$ by the waveguide transmission (50 $\%$) and by exp (-OD$_B$), where OD$_B$ is the background absorption in the transparency window, due to imperfect optical pumping (in our case OD$_B =1$, see dashed line in Fig. \ref{AFC}(b)). We emphasize that the waveguide and background losses are not fundamental and can be significantly reduced, by improving mode matching and by using shorter waveguides and/or optimized optical pumping techniques, respectively \cite{Hedges2010}.

When a control pulse is applied before the rephasing of the atomic excitations  the AFC echo is partially suppressed as the transfer to the spin state takes place (see Fig.  \ref{SWS}(b)). After a controllable time  $T_s$, a second control pulse is applied and the spin-wave echo is retrieved (solid blue curve). The control pulses have a Gaussian waveform and are frequency chirped by $1.5 \,\mathrm{MHz}$.  To confirm that the input light field is stored as a spin-wave, we measure its decay when increasing the spin storage time $T_s$. Assuming a Gaussian decay, we extract an inhomogeneous spin-broadening of $\gamma_{inh} = (23.6 \pm 7.7) \,\mathrm{kHz}$, compatible with those evaluated in different spin-storage experiments in \PrYSO \, \cite{Afzelius2010, Gundogan2013, Gundogan2015}. We observe an echo up to a total storage time of $t_s=1/\Delta+T_s=15\, \mu s$, more than two orders of magnitude longer that previous AFC demonstrations (at the excited state) in waveguides \cite{Saglamyurek2014}. The maximal internal  spin-wave efficiency  is $\eta_{SW} =2\,\%$. Similarly to the storage in the excited state,  waveguide and background loss have to be taken into account for estimating the device efficiency. The transfer efficiency of the control pulses is $\eta_T = 50\,\%$ for a laser power $P_p = 375 \,\mu \mathrm{W}$. The transfer efficiency is estimated as $\eta_{SW} = \eta_{AFC} \times {\eta_{T}}^2 \times \eta_C$ where $\eta_C$ accounts for the decoherence in the spin state and is evaluated from $\eta_T = \frac{\eta_{SW}} {\eta_{SW} (0)}$, where  $\eta_{SW}(0)$ is the storage efficiency at $T_s = 0$. 
It is worth noting that the AFC and spin wave efficiencies achieved in the waveguide sample are comparable to those obtained in similar experiments performed with weak coherent states at the single photon level in a shorter ($3\,\mathrm{mm}$) \PrYSO \,crystal \cite{Gundogan2015}, but the required laser power of the control beam is more than 50 times lower. The possibility to operate coherent population transfer with significantly lower power is a fundamental advantage as the noise given by the control pulses is the main limitation for the implementation of many on-demand storage protocols at the single photon level. Moreover, the number of atoms addressed is also decreased by the same factor, which will contribute strongly to reduce the noise. Contrarily to the bulk case \cite{Gundogan2015}, the integrated design does not offer the possibility to spatially separate the control and echo modes (e.g. with an angle), if both propagate in the waveguide. However, a counter-propagating configuration could be adopted. Our result is thus very promising for the ultimate goal of extending the storage protocol to the quantum regime \cite{Gundogan2015, Jobez2015}. In addition, it is worth noting, that an improved light-matter interaction in confined structures is also beneficial in view of observing and exploiting non-linear effects at the single photon level \cite{Sinclair2015}.

\section{Conclusions}
In conclusion, we have demonstrated an optical memory based on laser written waveguides in a \PrYSO \, crystal. We showed that the waveguide fabrication does not alter the coherence properties of the bulk crystal, and that the light confinement in the crystal increases the  light-matter interaction (as measured by the Rabi frequency) by a factor of 6 compared to a bulk crystal with the same focusing. In addition, we reported a proof-of-principle experiment of light storage  using the AFC protocol. We stored strong light pulses both in the excited and ground states of Pr ions. The latter represents the first demonstration of an integrated on-demand spin-wave memory. These results show that integrated optical memories can be realized using laser written waveguides, a versatile and widely used  technique in integrated quantum photonics. This opens new perspectives for the realization of long-lived integrated spin-wave quantum memories.

\textbf{Acknowledgments.} We acknowledge financial support by the ERC starting grant QuLIMA, by the Spanish Ministry of Economy and Competitiveness (MINECO) and Fondo Europeo de Desarrollo Regional (FEDER) (FIS2012-37569), by MINECO Severo Ochoa through grant SEV-2015-0522 and the PhD fellowship program (for A.S.), by AGAUR via 2014 SGR 1554,  by  People Programme (Marie Curie Actions) of the EU FP7 under REA Grant Agreement No. 287252, by Fundaci\'o Privada Cellex, and by the  European project QWAD (FP7-ICT-2011-9-600838).

\end{document}